\documentclass[12pt]{article}
\usepackage{epsfig}

\def\be{\begin{equation}}
\def\ee{\end{equation}}
\def\l{\label}

\begin{document}

\begin{titlepage}
\title{Combinatorial interpretation of Haldane-Wu fractional exclusion statistics}
\author{A.K. Aringazin and M.I. Mazhitov\\
Department of Theoretical Physics, Institute for Basic Research,\\
Eurasian National University, Astana 473021 Kazakstan}

\date{15 February 2002}

\maketitle

\abstract{Assuming that the maximal allowed number of identical
particles in state is an integer parameter, $q$, we derive the
statistical weight and analyze the associated equation which
defines the statistical distribution. The derived distribution
covers Fermi-Dirac and Bose-Einstein ones in the particular cases
$q=1$ and $q \to \infty$ ($n_i/q \to 1$), respectively. We show
that the derived statistical weight provides a natural
combinatorial interpretation of Haldane-Wu fractional exclusion
statistics, and present exact solutions of the distribution
equation.\\[7pt] PACS 05.30Pr}

\end{titlepage}

\section{Introduction}

Statistics which are different from Fermi-Dirac and Bose-Einstein
ones become of much interest in various aspects. A recent example
is given by Haldane-Wu fractional exclusion statistics (FES)
\cite{Haldane, Wu94} which is used to describe elementary
excitations of a number of exactly solvable one-dimensional models
of strongly correlated systems, and other models \cite{Wu94, Wu}.
This statistics is based on the statistical weight, which is a
generalization of Yang-Yang \cite{Yang} state counting as
mentioned by Wu,
\be\l{Haldane} W_i =
\frac{(z_i+(n_i-1)(1-\lambda))!}{n_i!(z_i-\lambda n_i -
(1-\lambda))!},
\ee
where the parameter $\lambda$ varies from $\lambda=0$
(Bose-Einstein) to $\lambda=1$ (Fermi-Dirac). This formula is a
simple generalization and interpolation of Fermi and Bose
statistical weights. While there is no physical meaning ascribed
to $\lambda$ here, the physical interpretation of
Eq.~(\ref{Haldane}) is that the {\sl effective} number of
available single-particle states {\sl linearly} depends on the
number of particles,
\be\l{zH}
 z_i^{f} = z_i -(1-\lambda)(n_i-1), \quad
 z_i^{b} = z_i -\lambda(n_i-1),
\ee
for fermions and bosons, respectively. This is viewed as a
defining feature of the fractional exclusion statistics.

In the present paper, we show that the equation which defines
Haldane-Wu statistical distribution can be derived from a
different statistical weight, which has a clear combinatorial and
physical treatment. Also, we present exact solutions of this
equation.

\section{The combinatorics}

A number of quantum states of $n_i$ identical particles occupying
$z_i$ states, with up to $q$ particles in state, $1\leq q
\leq n_i$, can be counted as follows.

We consider a configuration defined as that it has a maximal
possible number of totally occupied states (exactly $q$ particles
in state). A number of such totally occupied states is an integer
part of $n_i/q$ which we denote by $\left[\frac{n_i}{q}\right]$.
If $q$ is a divisor of $n_i$ we have identically
$\left[\frac{n_i}{q}\right] = n_i/q$, so that the number of
unoccupied states is $z_i-\frac{n_i}{q}$. If $q$ is not a divisor
of $n_i$ we have one partially occupied state, so that the number
of unoccupied states is $z_i-\frac{n_i}{q}-1$. We write a combined
formula of the statistical weight for both the cases as
\be\l{Wi}
W_i = \frac{\left(z_i+n_i-\left[\frac{n_i}{q}\right] \right)!}
{n_i!\left(z_i-\left[\frac{n_i}{q}\right]-l\right)!},
\ee
where $l=0$ or 1 if $n_i/q$ is integer or noninteger,
respectively; $i=1,2,\dots, m$.

In the particular cases, $q=1$ and $q=n_i$, we have
$\left[\frac{n_i}{q}\right] = n_i/q$ and $l=0$ so that
Eq.~(\ref{Wi}) reduces to Fermi-Dirac and Bose-Einstein
statistical weights, respectively,
\be\l{FDBE}
 W_i =\frac{z_i!}{n_i!(z_i-n_i)!},
 \quad
 W_i =\frac{(z_i+n_i-1)!}{n!(z_i-1)!}.
\ee

As one can see, the effective number of available single-particle
states derived from Eq. (\ref{Wi}),
\be\l{zAM}
 z_i^{f} = z_i -\left(1-\frac{1}{q}\right)n_i, \quad
 z_i^{b} = z_i -\frac{1}{q}n_i+1,
\ee
for fermions and bosons, respectively, is linear in $n_i$. With
the identification of the parameters, $1/q=\lambda$, and the
redefinition, $z_i \to z_i - (1-\lambda)$,  the statistical weight
(\ref{Wi}) coincides with Haldane-Wu statistical weight
(\ref{Haldane}), for the case of integer $n_i/q$. Consequently,
the obtained statistical weight (\ref{Wi}) corresponds to a kind
of fractional exclusion statistics. To verify whether (\ref{Wi})
leads to Haldane-Wu distribution we obtain below the equation
which governs statistical distribution.

\section{The distribution function}

Starting with Eq.~(\ref{Wi}), we follow usual technique of
statistical mechanics to derive the associated most-probable
distribution of $n_i$.

The thermodynamical probability is $W=\prod W_i$, and the entropy,
$S=k\ln W$, can be calculated by using the approximation of big
number of particles, $n! \simeq n^n e^{-n}$ for big $n$. Assuming
conservation of the total number of particles, $N=\sum n_i$ and
the total energy, $E=\sum n_i\varepsilon_i$, variational study of
$S$ corresponding to an equilibrium state gives us
\begin{eqnarray}\l{S}
 \delta S = k\sum\limits_i\biggl[
 \left(1- \frac{1}{q}\right)\ln\left(n_i +z_i
 -\frac{n_i}{q}\right)
 - \ln n_i  \nonumber \\
 + \frac{1}{q}\ln\left(z_i - \frac{n_i}{q}\right)
  - \alpha - \beta\varepsilon_i
 \biggr]\delta n_i =0,
\end{eqnarray}
where $\alpha$ and $\beta$ are Lagrange multipliers, and we have
used $\left[\frac{n_i}{q}\right] \simeq \frac{n_i}{q}$ and $l=0$
for big $n_i$. Using the notation $\kappa = 1/q$ and inserting
$\alpha=-\mu/kT$ and $\beta = 1/kT$ (obtained via an
identification of $S$, at $q=1$, with the thermodynamical
expression), we rewrite Eq.~(\ref{S}) as
\be\l{AM}
\frac{(z_i+(1-\kappa)n_i)^{1-\kappa}(z_i-\kappa n_i)^\kappa}{n_i}
= \exp\frac{\varepsilon_i-\mu}{kT}, \quad \kappa= 1, \frac{1}{2},
\frac{1}{3}, \dots
\ee

To draw parallels with Haldane-Wu statistics below we make
analytic continuation of the discrete parameter $\kappa$ assuming
$\kappa \in [0,1]$. Under this condition, the derived distribution
equation (\ref{AM}) {\sl does reproduce} that of Haldane-Wu
fractional exclusion statistics (Eq.~(14) of ref. \cite{Wu94}),
with $\kappa=\lambda$.

Below, we turn to consideration of properties and exact solutions
of Eq.~(\ref{AM}).

In general, Eq.~(\ref{AM}) can not be solved exactly with respect
to $n_i$. However, for $\kappa=1$ and $\kappa \to 0$ ($\kappa n_i
\to 1$), it becomes linear in $n_i$ and gives Fermi and Bose
distributions, respectively. Also, we note that for $\kappa=$ 1/2,
1/3, and 1/4 the equation contains a polynomial of degree up to 4
so that it can be solved exactly for all these cases.

A convenient expression for $n_i$ obeying Eq.~(\ref{AM}) is given
by \cite{Wu94}
\be\l{solution}
n_i = \frac{1}{w(x)+\kappa},
\ee
where we have redefined, $n_i/z_i\to n_i$,
$x\equiv\exp[(\varepsilon_i-\mu)/kT]$, and the function $w(x)$
satisfies
\be\l{w}
(1+w)^{(1-\kappa)}w^\kappa = x.
\ee

Remarkably, exclusons which are "close" to fermions can be
described in terms of exclusons which are "close" to bosons. In
fact, we note that Eq.~(\ref{AM}) is invariant under a set of
transformations,
\be\l{symm}
\kappa \to 1- \kappa, \quad n_i \to -n_i, \quad x\to -x,
\ee
for $\kappa \not=0, 1$. Therefore, if $n_i(x,\kappa)$ satisfies
Eq.~(\ref{AM}) then the function $m_i = -n_i(-x,1-\kappa)$
satisfies the same equation. Thus, we obtain the following general
relation
\be\l{relat}
n_i(-x,1-\kappa) = -n_i(x,\kappa), \quad \kappa\not=0,1.
\ee
We see that the distribution $n_i$ of exclusons for, e.g.,
$\kappa=1/200 \simeq 0$ can be obtained from that of "dual"
exclusons, with $\kappa = 1-1/200=199/200 \simeq 1$.

The values $\kappa=1$ and $\kappa \to 0$ ($\kappa n_i \to 1$) are
the only two points of {\sl degeneration} of Eq.~(\ref{AM}).
Hence, any "deviation" from Fermi or Bose statistics is
characterized by a sharp change of statistical properties, sending
us to consideration of exclusons. Consequently, we can divide
particles into three main types, genuine fermions, genuine bosons,
and exclusons ($\kappa \in ]0,1[$), since their statistical
distributions obey {\sl different non-degenerate} equations.

A fixed point of the map $\kappa \to 1-\kappa$ is $\kappa=1/2$.
Hence it represents a special case worth to be considered
separately. In this case, Eq.~(\ref{AM}) allows an exact solution
and the result is (positive root) \cite{Wu94}
\be\l{semion}
n_i =  \frac{2}{\sqrt{1+4x^2}}
=\frac{2}{\left(1+4\exp\frac{2(\varepsilon_i-\mu)}{kT}
\right)^{1/2}}.
\ee
This distribution represents statistics with up to two particles
in state, $q=2$ (semions).

We have obtained exact solutions (real roots) of Eq.~(\ref{AM})
for $\kappa=1/3$ and 2/3 which we write as
\be\l{triplon}
 n_i = \frac{3}{f + f^{-1} \mp 1}, \quad
 \ee
where
\be\l{ys}
 f=\left[2\sqrt{y(y \mp 1)} + 2y \mp 1\right]^{1/3}, \quad
 y=2\left(\frac{3x}{2}\right)^3 \pm 1.
\ee
From Eqs.~(\ref{triplon}) and (\ref{ys}) one can see how exclusons
with $\kappa=1/3$ (upper sign) are related to exclusons with
$\kappa=2/3$ (lower sign) that agrees with Eq.~(\ref{relat}).
Also, for $\kappa=1/4$ and 3/4 we have obtained the following
exact solutions (positive real roots):
\be\l{quadron}
n_i= \frac{4}{\sqrt{ 2g^{-1/2} - g + 3} \pm g^{1/2} \mp 2},
\ee
where
\be\l{fy}
g = \frac{3}{2}\left([z^2(z+2)]^{1/3} + [z(z+2)^2]^{1/3}\right)+1,
\quad
 z=\sqrt{3\left(\frac{4x}{3}\right)^4 + 1} -1.
\ee

Plots of $n_i(x)$ for various $\kappa$ are presented in Fig.~1,
from which one can see that these exclusons behave similar to
fermions.

\begin{figure}[ht]
\label{Fig1}
\begin{center}
\epsfxsize=\textwidth
\parbox{\epsfxsize}{\epsffile{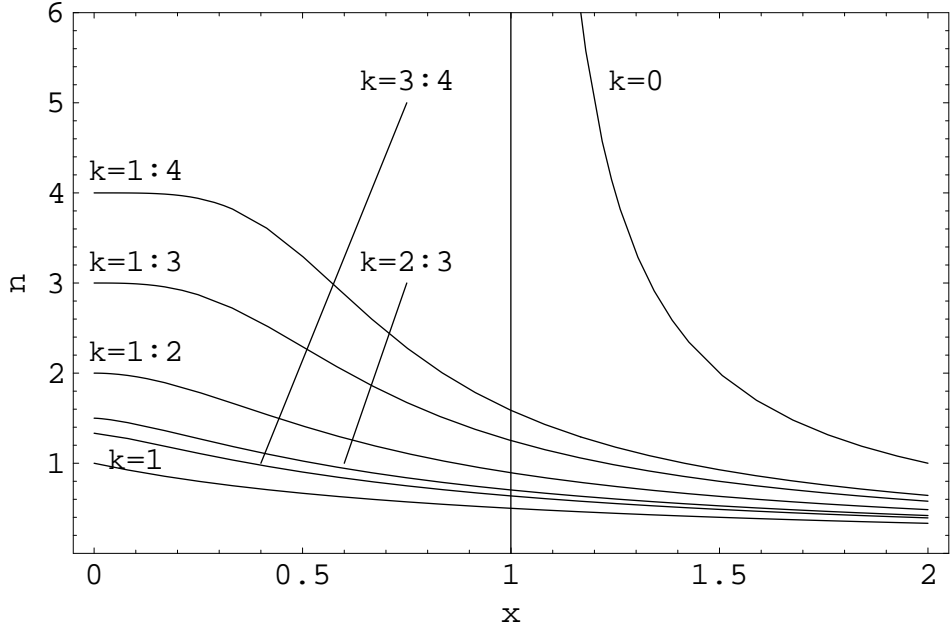}}
\end{center}
\caption{Statistical distribution $n_i$ as a function of
$x=\exp[(\varepsilon_i-\mu)/kT]$, for $\kappa=1$ (fermions),
$\kappa=0$ (bosons), $\kappa=1/2$ (semions, Eq.~(\ref{semion})),
$\kappa=1/3$ (Eq.~(\ref{triplon}), upper sign), $\kappa=1/4$
(Eq.~(\ref{quadron}), upper sign), $\kappa=2/3$
(Eq.~(\ref{triplon}), lower sign), and $\kappa=3/4$
(Eq.~(\ref{quadron}), lower sign).}
\end{figure}

Distributions of exclusons can be obtained from a different
approach, based on the canonical statistical sum which implies the
mean number of particles,
\be\l{sum}
n = \frac{\sum\limits_{N=0}^{q}N
x^{-N}}{\sum\limits_{N=0}^{q}x^{-N}}.
\ee
This formula gives (exact) Fermi and Bose distributions for $q=1$
and $q\to\infty$, respectively, while for arbitrary $q \geq 1$ the
sum is
\be\l{excluson}
n = -\frac{x^{1+q} -(1+q)x +q}{(x^{1+q}-1)(x-1)}, \quad
q=1/\kappa.
\ee
In Fig.~2, we compare distributions (\ref{excluson}) with exact
solutions shown in Fig.~1. One can see that deviations become
considerable as $\kappa$ goes to smaller values. However, we
expect that near $\kappa=0$ there should be a better
correspondence since one approaches the other interpolation
endpoint (bosons). We treat (\ref{excluson}) as an approximate
result which is useful since it gives a single simple distribution
formula for all exclusons, $\kappa\in [0,1]$.

A connection between the two approaches requires a deeper study
which can be made elsewhere.

\begin{figure}[ht]
\label{Fig1}
\begin{center}
\epsfxsize=\textwidth
\parbox{\epsfxsize}{\epsffile{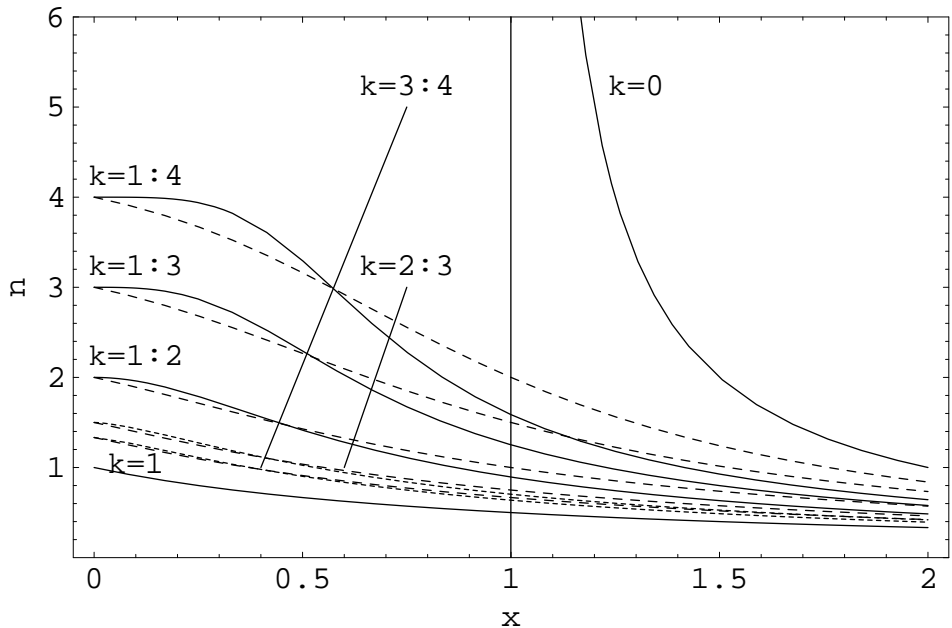}}
\end{center}
\caption{Statistical distribution $n_i$ as a function of
$x=\exp[(\varepsilon_i-\mu)/kT]$, for various $\kappa$; dashed
lines represent the approximation (\ref{excluson}) to exact
solutions (solid lines) shown in Fig.~1.}
\end{figure}

\section{Conclusions}

\indent

 (i) The derived statistical weight (\ref{Wi}) and
Haldane-Wu statistical weight (\ref{Haldane}) lead to the same
distribution equation (\ref{AM});

(ii) Haldane-Wu parameter $\lambda$ acquires a physical meaning of
an inverse of the maximal allowed occupation number in state,
$\lambda = 1/q$, similar to the inverse of the statistical factor
as shown by Wu \cite{Wu94};

(iii) Within fractional exclusion statistics, the generalized
Pauli exclusion principle reads that a maximal allowed occupation
number of identical particles in state is an integer,
$q=1,2,3,\dots$, i.e. $n_i/z_i \leq 1/\lambda$ as formulated by
Wu~\cite{Wu94}. We stress that in our approach we use this
principle as a basis to calculate statistical weight (\ref{Wi})
rather than derive it {\it a posteriori} from the analysis of a
statistical weight or distribution function;

(iv) While Haldane-Wu parameter $\lambda$ is assumed to vary
continuously, the statistical parameter $\kappa =1/q$ runs over
{\sl discrete} set of values, $\kappa = 1, 1/2, 1/3, \dots$ This
may be an important difference since physically acceptable
solutions of Eq.~(\ref{AM}) may not exist for all values of
$\kappa \in ]0,1[$, while $\kappa = 1, 1/2, 1/3, \dots$ guarantees
a polynomial structure of Eq.~(\ref{AM}), with physically
acceptable solutions;

(v) The equation (\ref{AM}), which defines statistical
distribution of exclusons, $\kappa \in ]0,1[$, has a remarkable
symmetry (\ref{symm}) which allows to interconnect solutions $n_i$
for $\kappa$ and $1-\kappa$ due to Eq.~(\ref{relat}).

In summary, we have shown that Haldane-Wu fractional exclusion
statistics finds a natural combinatorial and physical
interpretation in accord to Eq.~(\ref{Wi}), and presented exact
solutions of Eq.~(\ref{AM}).

\end{document}